\RequirePackage{fixltx2e}
\documentclass[preprint, floatfix, superscriptaddress]{revtex4-1}
\usepackage{hyperref}
\usepackage{graphicx,graphics}
\usepackage{amssymb,amsfonts}
\usepackage{amsmath}
\usepackage{siunitx}
\usepackage[T1]{fontenc}
\usepackage{lmodern}

\begin{document}
\def\mean#1{\left< #1 \right>}
\newcommand\solidrule[1][0.4cm]{\rule[0.5ex]{#1}{.8pt}}
\newcommand\dashedrule{\mbox{%
		\solidrule[0.8mm]\hspace{0.6mm}\solidrule[0.8mm]\hspace{0.6mm}\solidrule[0.8mm]}}

\title{Insights into the Hierarchical Structure of Spider Dragline Silk Fibers: Evidence for Fractal Clustering of $\beta$-Sheet Nano-Crystallites}
\author{Qiushi Mou}
\affiliation{Department of Physics, Arizona State University, Tempe, AZ 85281, USA}
\author{Chris J. Benmore}
\affiliation{X-Ray Science Division, Advanced Photon Source, Argonne National Laboratory, 9700 S. Case Avenue, Illinois 60439, USA}
\author{Warner S. Weber}
\affiliation{Department of Chemistry \& Biochemistry, Arizona State University, Tempe, AZ 85281, USA}
\author{Jeffery L. Yarger}
\email{jyarger@gmail.com}
\affiliation{Department of Physics, Arizona State University, Tempe, AZ 85281, USA}
\affiliation{Department of Chemistry \& Biochemistry, Arizona State University, Tempe, AZ 85281, USA}

\begin{abstract}
Spider dragline silk is one of the toughest materials known and understanding the hierarchical structure is a critical component in the efforts to connect structure to function.  In this paper, we take the first step in elucidating the hierarchical fractal structure of $\beta$-sheet nano-crystallites, which form a robust self-similar network exhibiting an non-linear mechanical property.  A combined small angle X-ray scattering (SAXS) and wide-angle X-ray scattering (WAXS) study of the nano-crystalline component in dragline silk fibers from several species of spiders including, \textit{Latrodectus hesperus}, \textit{Nephila clavipes}, \textit{Argiope aurantia} and \textit{Araneus gemmoides} is presented. SAXS structure factors exhibit a `lamellar peak' in the q-range from 0.60 to 0.82 nm\textsuperscript{-1} for various spider dragline silk fibers, indicating the presence of strong nano-crystal ordering on the $>$10 nm length-scale. The stochastically reconstructed electron density maps indicate that the $\beta$-sheet crystals are hierarchically structured as mass fractals and that nano-crystals tend to form 10 to 50 nm sized clusters with long-range crystalline ordering. This nano-crystal ordering along the fiber axis also helps to explain the difference between axial and radial sound velocities recently measured by Brillouin spectroscopy.
\end{abstract}
\maketitle

\section{Introduction}
Spider dragline silk fibers have excellent reversible extensibility and high tensile strength \cite{romer, cr010194g}. Extensive studies on the molecular structure of spider silk have been performed over the years and it's widely believed that oriented anti-parallel $\beta$-sheet nano-crystals are the key contributor to spider silk's excellent mechanical properties \cite{nmat2704}. The molecular structure and chemical composition of spider silks have been studied extensively by solid state nuclear magnetic resonance spectroscopy and these studies have shown that a large fraction of the amino acid sequences are poly-(Gly-Ala) and poly-Ala repeats \cite{bm400791u, C3SM52187G, Hayashi1999271, Xu01091990}, which form rigid anti-parallel $\beta$-sheet nano-crystals through the periodic hydrogen bond assemblies \cite{nmat2704,PhysRevLett.100.198301, Keten02062010}. The less ordered amino acids are in the form of random-coil like helical secondary structures in which the $\beta$-sheet nano-crystals are embedded. The elastic and random-coil like $\alpha$- and 3\textsubscript{10}- helical structures are abundant and occupy a large volume of the fiber body, acting as the interconnections among the rigid crystals \cite{C3CC43737J}. The crystal structure and physical size of individual $\beta$-sheet nano-crystal has also been resolved by X-ray diffraction studies. Past WAXS studies, as well as this work, have confirmed that typical $\beta$-sheet crystals have an orthorhombic unit cell with their physical sizes ranging from 2 to 4 nm, when produced at the natural extrusion speed \cite{Riekel2001203, C2SM25373A}.

While the nano-scale dimensions of the $\beta$-sheet crystals have been well studied, their hierarchical structures and relation to the macroscopic mechanical properties still hasn't been solved. Being the most rigid objects in the spider dragline silk, the $\beta$-sheet crystals play a crucial role in determining its mechanical properties. If we assume the $\beta$-sheet crystals to be the building blocks of the cylindrical-shape fiber, then its physical size, inter-crystal distances and long-range packing pattern will be the key parameters that define the macroscopic mechanical properties. Fortunately, several microscopy studies have gained insight on the crystallite structures of the spider silk fibers. Scanning electron microscopy (SEM) studies have shown that the texture of ion-etched silk fiber's is rather rough, as scattered crystalline-rich regions about the size of 20 nm to 50 nm have been observed across the silk fibril \cite{kitagawa}. Transmission electron microscopy (TEM) image shows relatively large crystallites on the scale of 70 to 120 nm are embedded in the amorphous matrix \cite{JMI947}. To date, there is still no consistent model to describe the hierarchical structure of these large crystalline regions, as the dominant scattering centers, i.e. $\beta$-sheet crystals, only span several nanometers in all three dimensions. By analyzing of the SAXS structure factor using a stochastic reconstruction method, we provide evidence that the crystalline structure of spider dragline silk fiber is mass fractal, accompanied by dense clustered packing of the nano-crystals. The `large crystals' that span up to 70 nm are composed of highly oriented and closely interlinked $\beta$-sheet crystals. In this study, we will present a combined WAXS and SAXS analysis of the crystalline phase in spider silks followed by the modeling of the spacial packing of the $\beta$-sheet crystals that reveals the underlying morphology of the crystalline structure within the spider dragline silk fibers.

\begin{table*}[hbt]
	\caption{Lattice parameters, nano-crystal sizes and inter-crystallite distance.}
	\label{tab:waxs}
	\begin{tabular*}{\hsize}
		{@{\extracolsep{\fill}}crrrrrrrrrrrc}
		\multicolumn{4}{l}{Lattice parameters}&
		\multicolumn{3}{l}{From (200)}&
		\multicolumn{3}{l}{From (120)}&
		\multicolumn{3}{l}{lamellar peak d-spacing}\cr
		\hline
		\multicolumn1c{Fiber}&\multicolumn1c{$a (\si{\angstrom})$}&
		\multicolumn1c{$b (\si{\angstrom})$}&\multicolumn1c{$c (\si{\angstrom})$}&
		\multicolumn1c{$\sigma$}&\multicolumn1c{$2\theta$}&
		\multicolumn1c{$\tau_{1} (\si{\angstrom})$}&
		\multicolumn1c{$\sigma$}&\multicolumn1c{$2\theta$}&
		\multicolumn1c{$\tau_{2} (\si{\angstrom})$}&\multicolumn1c{$d (\si{\angstrom})$}&
		\multicolumn1c{$d/\tau_{1}$}& \multicolumn1c{$d/\tau_{2}$}\cr
		\hline
		\textit{L. hesperus} &10.46 &9.62 &6.88    &1.102 &15.18 &21.3    &0.824 &18.24 &26.4  &77.1 &3.6 &2.9\cr
		\textit{N. clavipes} &10.70 &9.73 &6.86    &1.073 &15.21 &20.3    &0.960 &18.26 &22.7  &83.2 &4.1 &3.6\cr
		\textit{A. aurantia} &10.52 &9.74 &6.92    &1.107 &15.06 &19.6    &0.933 &18.13 &23.4  &108.7 &5.5 &4.6\cr
		\textit{A. gemmoides} &10.58 &9.67 &7.09    &1.023 &15.09 &21.3    &0.902 &18.10 &23.7  &106.6 &5.0 &4.5\cr
		\hline
	\end{tabular*}
\end{table*}

\begin{figure}
	\includegraphics[width=.6\textwidth]{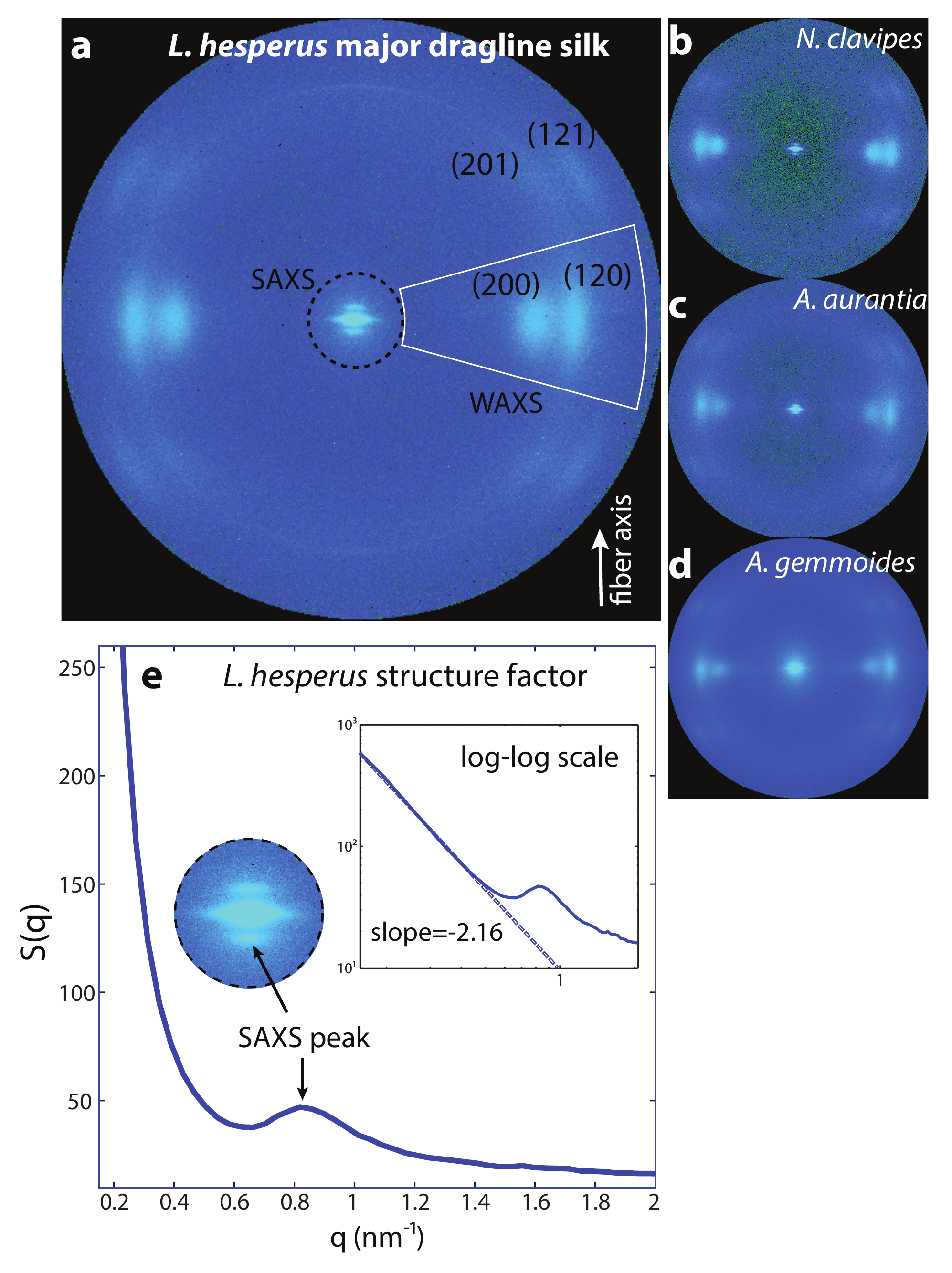}
	\caption{\textbf{A combined small angle X-ray scattering (SAXS) and wide-angle X-ray scattering (WAXS) image.} (\textbf{a}) WAXS pattern of \textit{L. hesperus} (Black Widow) major ampullate (dragline) silk. The wide-angle diffraction spots have been assigned with Miller indices, from which the unit cell parameters have been calculated using an orthorhombic crystal model. Gaussian peak fitting was applied to the wedge shaped region (white line) and the Scherrer equation was used to determine the average nano-crystallite dimensions. The center region (q<1.3 nm\textsuperscript{-1}) corresponds to the SAXS scattering pattern. (\textbf{b-d}) Samples of \textit{N. clavipes}, \textit{A. aurantia} and \textit{A. gemmoides} show similar wide-angle scattering patterns. (\textbf{e}) Azimuthal integration of scattering intensity from \textit{L. hesperus} dragline silk fibers. The lamellar peak is located at q=$0.82\pm 0.01$ nm\textsuperscript{-1}. The inset shows the SAXS structure factor $S(q)$ on log-log scale, where the `matrix knee' represents the intermediate length scale (1 nm to 200 nm) and exhibits linearity with a slope of 2.16.}
	\label{fig:xray}
\end{figure}

\section{Results and discussion}
Fig.1 shows the WAXS and azimuthal integrated SAXS profile of \textit{L. hesperus}  (Black Widow) dragline silk fiber. The diffraction pattern is divided into two distinct regions: the center small-angle region (q<1.3 nm\textsuperscript{-1}) and the outer wide-angle region. As shown in Fig.1a, the diffraction spots have been assigned with Miller indices from which we have calculated the unit cell parameters basing on an orthorhombic unit cell model \cite{C2SM25373A}. To retrieve the crystal sizes from X-ray data, we integrated the wedge shape equatorial region containing reflections (200) and (120) (Fig.\ref{fig:xray}a), and then applied a 1-D Gaussian peak fitting procedure [see Fig.S1]. The full maximum at half width (FMHW) of the fitted Gaussian peaks are evaluated using the Scherrer equation \cite{PhysRev.56.978}
\begin{equation}
	\tau = \frac{K\lambda}{\beta\cos \theta}
\end{equation}
where $K=0.9$, $\lambda=\SI{1.38}{\angstrom}$, $\beta$ is the FWHM of fitted peak and $\theta$ is the Bragg angle, to calculate the nano-crystal physical dimension $\tau$. The orthorhombic unit cell and crystal sizes are summarized in Table.\ref{tab:waxs}. The crystal size in all three dimensions (a, b, c) shows very small variations for the four species of dragline silk examined here. The crystal sizes have a narrow distribution, namely from 19.6 to 21.3 \si{\angstrom} for that calculated from (200) reflection and from 22.7 to 26.4 \si{\angstrom} for the (120) reflection. Although different species of dragline silk shows significant differences in alanine content and mechanical properties\cite{bm400791u}, their building blocks are surprisingly similar in terms of physical appearance. Fig.1e shows the small-angle scattering profile of \textit{L. hesperus} (Black Widow) major ampullate (dragline) silk fibers, where the characteristic lamellar peak is located at 0.82 nm\textsuperscript{-1}. The characteristic correlation peaks between 0.5 and 1.2 nm\textsuperscript{-1} have been observed frequently in polymer and copolymer materials \cite{ma00246a013}. The correlation peaks manifest certain long-range ordering of the crystalline phases and such ordered state is crucial to the functional behavior of the material, such as water channels morphology observed in nafion polymers \cite{Schmidt-Rohr:nmat}. Therefore, we propose that it's the intermediate length-scale morphology of the $\beta$-sheet crystals that determines the macroscopic properties such as mechanical strength, elasticity and thermal conductivity. 

\begin{figure}[htb]
	\centering
	\includegraphics[width=.55\textwidth]{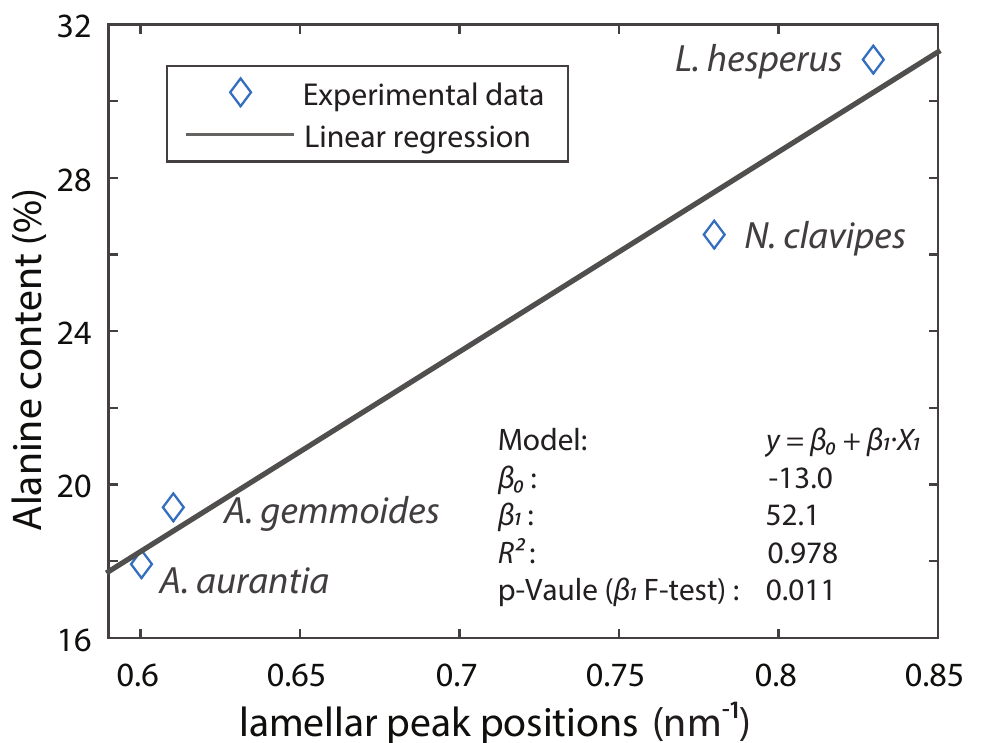}
	\caption{\textbf{Correlation of lamellar peak positions to Alanine content.} Alanine in spider silks is shown to primarily occur in $\beta$-sheet nano-crystals and an increase in alanine content in the spider silk protein is a contributing factor to increased crystallinity associated with the nanoscale clusters which give rise to the lamellar peaks. The alanine content data was previously retrieved from solid-state nuclear magnetic resonance experiments.}
	\label{fig:linear}
\end{figure}

Since $\beta$-sheet crystals are identified as the only kind of strong scattering centers in spider dragline silks, it's natural to assume $\beta$-sheet crystals as the origin of the SAXS lamellar peaks. The packing pattern of the crystals has unique intermediate-range ordering, which leads to the appearance of strong lamellar peaks in the $q<1$ nm\textsuperscript{-1} range of the SAXS structure factor \cite{Schmidt-Rohr:ce5003,Pedersen:gl0327}. For strong correlation peaks to appear, it's important that the scattering centers maintain a closest approach distance $R_{CA}$ between each pair, as has been discussed by Yarusso and Copper \cite{ma00246a013}. In real space, the closest approach distance $R_{CA}$ limits the average spacing between each adjacent pair of scattering centers that reside in the amorphous backbone of biopolymer. Reflected in reciprocal space, the combination of closest approach distance and long range ordering will generate correlation peak in the small-angle scattering regime. For different species of spider dragline silks examined here, all of them exhibit correlation peaks in the range of 0.6 to 0.9 nm\textsuperscript{-1}. This indicates that the $\beta$-sheet crystals are not adjacent to each other statistically, but rather maintain an average closest distance among them, which can be quantified by the characteristic of SAXS lamellar peak. The SAXS signal exhibits an elliptical streak, elongated along the meridian direction. This indicates that the $\beta$-sheet nano-crystals have a rectangular shape with long axis parallel with fiber axis, as illustrated by Fig.S2 in supplementary material. Combining these information, we constructed a 2-dimensional electron density map in which the beta-sheet crystals are initialized with constrained orthorhombic geometry. The d-spacings of the lamellar peaks were taken as the average inter-crystal distance in constructing the initial model. This in turns determines the density of the $\beta$-sheet crystals in each silk species. The linear correlation between the lamellar peak position and alanine content further strengthens this assumption. 

Past solid state NMR experiments have shown that the fraction of alanine content, which primarily occur in the $\beta$-sheets, vary significantly among different species \cite{bm400791u, doi:10.1021/bm100399x}. By cross comparing the data from x-ray scattering and NMR, we have found the alanine content to be linearly correlated to lamellar peak's position in $q$-space, as shown in Fig.\ref{fig:linear}. With a relatively small inter-crystal distance, such as 83.2 and 77.1 \si{\angstrom} measured from \textit{N. clavipes} and \textit{L. hesperus} respectively, we find that the $\beta$-sheet crystals are more densely packed and thus leading to a higher fraction of beta-sheet crystals than the other two samples, namely \textit{A. aurantia} and \textit{A. gemmoides} (Fig.S3). 

\begin{figure}[htb]
	\includegraphics[width=.7\textwidth]{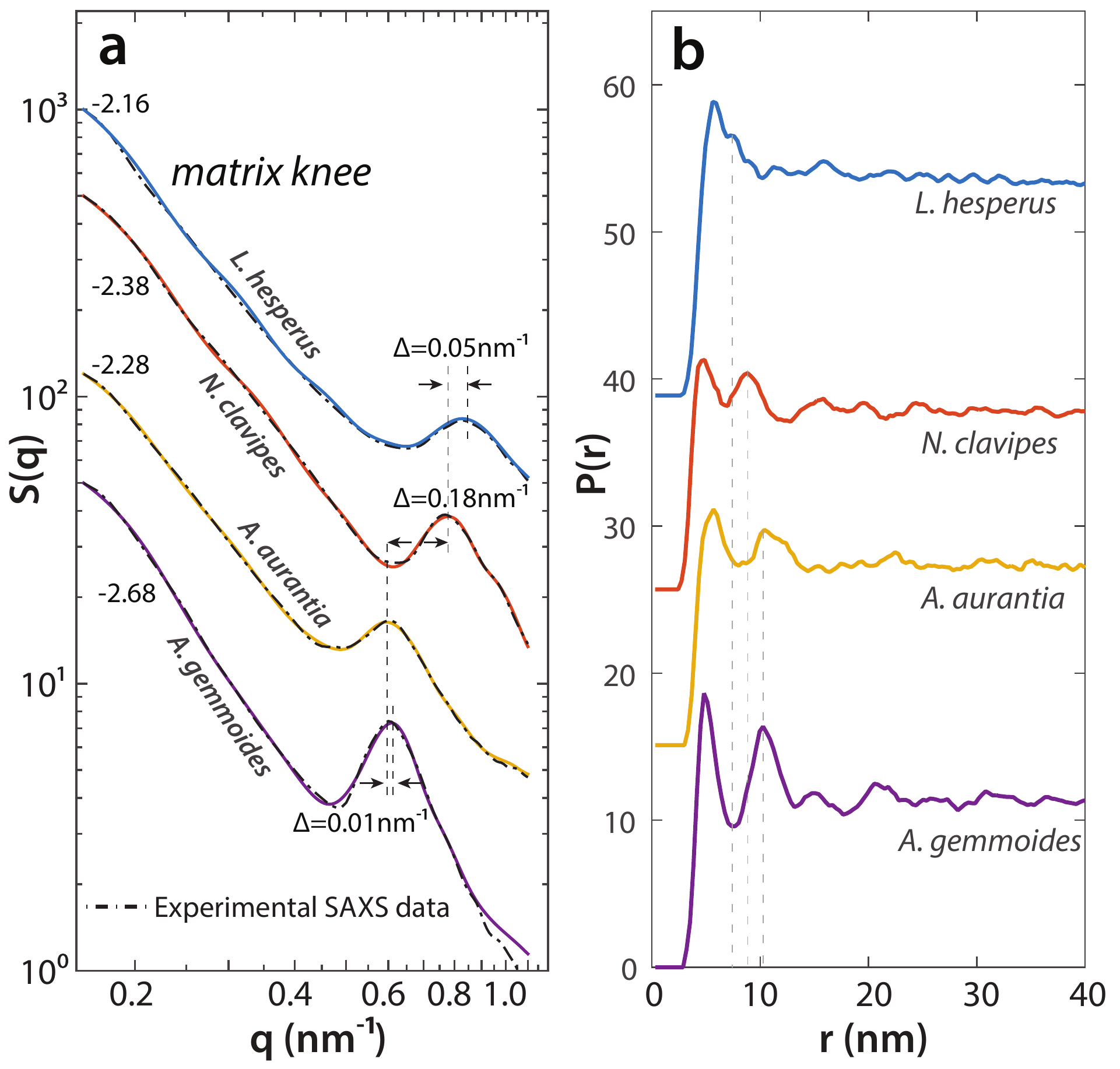}
	\caption{\textbf{Structure factors $S(q)$ and pair correlation function from reconstructed structure.} (\textbf{a}) The numerically simulated structure factor $S(q)$ (\solidrule) fits the experimental SAXS $S(q)$ (\dashedrule). The small-angle lamellar peaks' positions, from \textit{L. hesperus} (top) to \textit{A. gemmoides} (bottom),  are correspondingly located at 0.83, 0.78, 0.60 and 0.61 nm\textsuperscript{-1} with errors of approximately $\pm 0.01$nm\textsuperscript{-1}. The low-$q$ region of the structure factor $S(q)$, i.e. the matrix knees, exhibit linearity in all cases and the slopes range from -2.1 to -2.7 on the log-log scale. (\textbf{b}) Pair correlation function $P(r)$ calculated from the reconstructed electron density maps. The correlation function $P(r)$ were calculated from a population of around 5000 crystals and the curves were numerically smoothed. The intermediate range crystalline ordering is reflected as the multiple correlation peaks observed on the $P(r)$ curves in the range of 7 to 40 nm. }
	\label{fig:saxs}
\end{figure}

\begin{figure}[htb]
	\includegraphics[width=.7\textwidth]{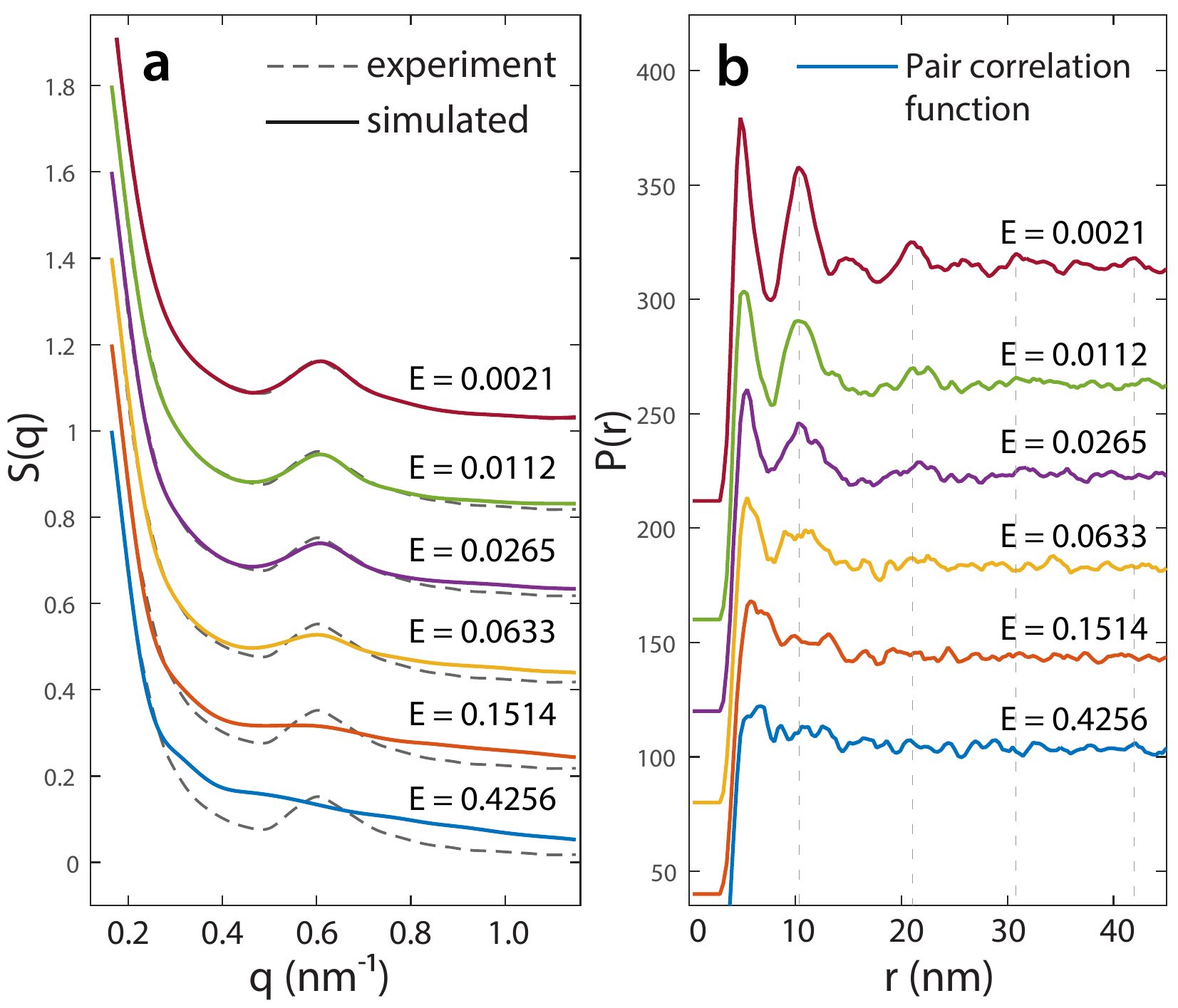}
	\caption{\textbf{Time evolution of the structure factor and the pair-correlation function $P(r)$ during the stochastic reconstruction on silk species \textit{A. gemmoides} .} (\textbf{a}) From bottom to top, as the stimulated annealing temperature $T$ drops, the calculated structure factor (\solidrule) converges to the experimental structure factor (\dashedrule), reducing the pseudo-energy $E$ as defined in Eq.\ref*{eq:engy}. (\textbf{b}) Initially at $E=0.4256$, the $P(r)$ function is absent from any correlation peak. As the stimulated annealing algorithm proceeded, the model build up intermediate range crystalline ordering which was reflected by the correlation peaks at 12, 15, 21, 26, 32, 37 and 42 nm on the top curve ($E=0.0021$) . The correlation function $P(r)$ is sampled from a model contains 4721 $\beta$-sheet crystals. }
	\label{fig:dyn}
\end{figure}

\begin{figure*}[hbt]
	\includegraphics[width=1.0\textwidth]{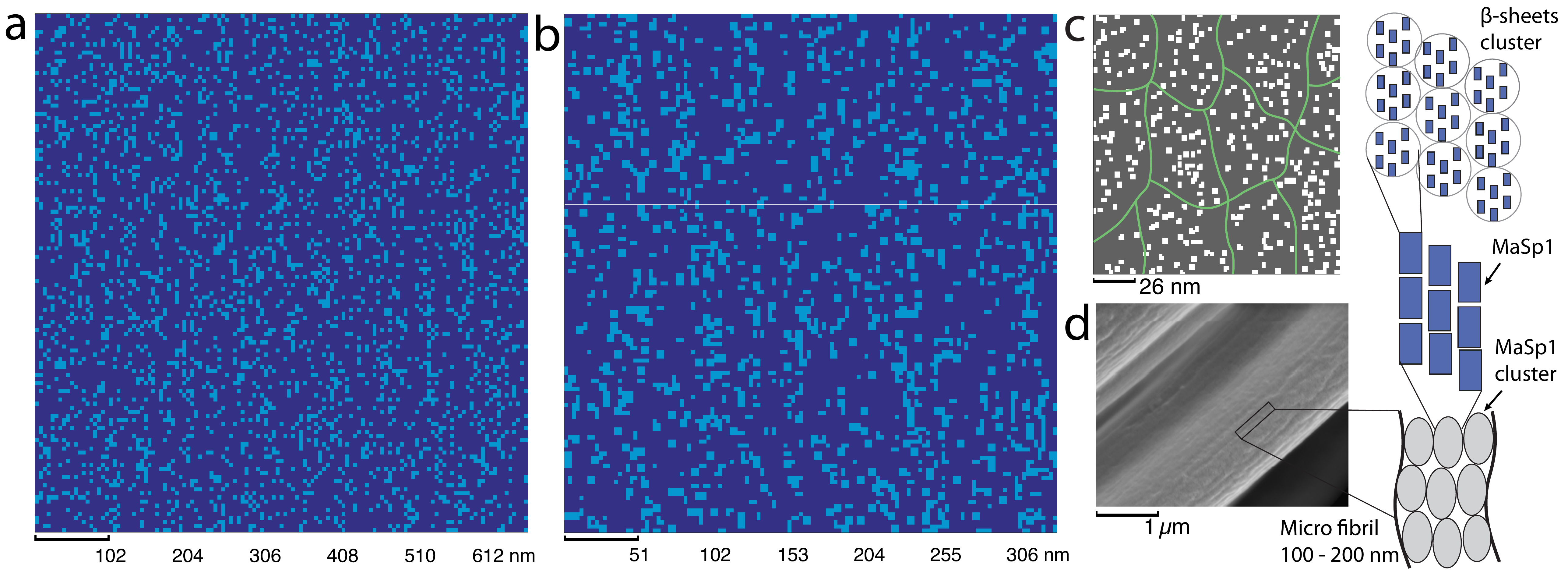}
	\caption{\textbf{Reconstructed electron density map of silk species \textit{A. gemmoides}}. (\textbf{a-c}) Coarse grained electron density map from high to low degree of coarse graining. Each map shows the magnified upper right quater of the previous. The texture on these three maps with drastically different length scales show self-similarity property. The clustering effect of $\beta$-sheets is visually emphasized by the boundaries in (c). (\textbf{d}) The schematic illustration of the hierarchical mass fractal structure of the $\beta$-sheet clusters in silk fiber, from SEM image to the basic $\beta$-sheets network. The micro-fibril (100-200 nm) is composed of functional mechanical units, and each unit (40-60 nm) is composed of multiple crystallites-rich MaSp1 proteins (10-20 nm). Within a MaSp1, it contains multiple $\beta$-sheet nano-crystals (2-3 nm). Such fractal structure is mechanical robust and exhibits non-linear force-extension behavior.}
	\label{fig:map}
\end{figure*}

The experimental and numerically simulated SAXS structure factors $S(q)$ are shown in Fig.\ref{fig:saxs}a. The lamellar peaks are present in all fiber samples and they range from 0.6 to 0.9 nm\textsuperscript{-1} with variations in the peak intensity. The lamellar peaks of \textit{A. aurantia} and \textit{A. gemmoides} are relatively close in q-space with a minuscule difference $\Delta q$=0.01 nm\textsuperscript{-1}, while \textit{L. hesperus} and \textit{N. clavipes} are slightly further apart with a difference of $\Delta q$=0.05 nm\textsuperscript{-1}. The structure factor curves are very well reconstructed across the entire collected q-range by the stimulated annealing reconstruction method. The lower bound of the $S(q)$ is limited by the detector coverage on 14-ID-B beam line while the upper bound is cut at q=1.2 nm\textsuperscript{-1}, beyond which point the structure factor $S(q)$ begins to exhibit WAXS feature. Fig.\ref{fig:saxs}b shows the pair correlation functions $P(r)$ calculated from the reconstructed electron density maps. The inter-molecular $\beta$-sheet pair correlation function $P(r)$ \cite{PhysRevB.80.024118,Proffen:gl0603} is defined as
\begin{equation}
	P(r) = \frac{1}{2\pi r\rho_{0}}\frac{1}{N}\sum\limits_{i=1}\sum\limits_{j\neq i}\frac{w_{i}w_{j}}{\mean{w}^2}\delta(r-r_{ij})
\end{equation}
,where $2\pi r\rho_{0}$ is the density of states in 2-D isotropic, homogeneous system, $w_{i}$ is the weighting factor for the corresponding scattering center, and $N$ is the $\beta$-sheet crystal population. The intermediate range crystalline ordering is reflected as the multiple correlation peaks observed on the $P(r)$ curves in the range of 7 to 40 nm. The first correlation peaks appear near 6 nm and these peaks arise from the dominant closest pair-pair interaction of the $\beta$-sheet crystals. The second peaks are in the range of 7 to 10 nm, which arise from the intra-cluster interaction. As shown in Fig.\ref{fig:saxs}b, the second peaks is shifting right top-down, which follows the density and inter-crystal spacing constrain, though the exclusion region was reduced during the stochastic reconstruction to allow higher mobility (Fig.4). The intensity and sharpness of the lamellar peak reflects strength of pair correlations directly. While the \textit{A. gemmoides} and \textit{N. clavipes} have the stronger lamellar peak (Fig.\ref{fig:saxs}a), they also exhibit more correlation peaks in the $>10$ nm range of their $P(r)$ function. This indicates that a larger fraction of the $\beta$-sheets are in long range ordered state for these two silk species. 

Fig.\ref{fig:dyn} shows the evolution of the stimulated structure factor and the pair-correlation function calculated from the reconstructed crystalline model. The correlation function shows no feature in the early stage of the reconstruction, mainly due to the randomized landscape of the crystals and the relatively small population, which varies between 4000 to 7000. As the structure factor converges to the experimental value, the correlation peaks gradually build up, notably for the peaks at 10, 21, 31 and 42 nm, indicated by the dashed line on Fig.\ref{fig:dyn}b. On one hand, this dynamic proves that one can reconstruct the pair correlation function through reciprocal-space reconstruction. On the other hand, it consolidate the belief that the SAXS lamellar peaks, i.e. the correlation peaks between the q range of 0.6 to 0.8 nm\textsuperscript{-1}, arise from the intermediate length scale ordering of the $\beta$-sheet crystals.

The coarse-grained electron density maps are shown in Fig.\ref{fig:map}(a-c). The lighter area indicates the presence of a higher density of $\beta$-sheet crystals in that region whereas the darker area represents the amorphous backbone, which represents diffuse X-ray scattering. The $\beta$-sheet crystal distributions are initialized with a lamellar modulation when building the initial model. The modulation, which typically has a strip size of $N_{lamellar}=128$  equaling to a physical size of 30 - 50 nm, is designated to represent the micro-fibril structure observed from both SEM \cite{kitagawa} and AFM experiments \cite{Du20064528}. The existence of the parallel lamellar nano-fibrils could explain the discrepancy between the spider silks' axial and radial sound velocities reported by Koski et. al \cite{nmat3549}. Along the axial direction of fiber, these parallel lamellar fibrils contain a high density of $\beta$-sheet crystals with crystalline ordering up to 30 nm, which can be observed on Fig.\ref{fig:map}b. The highly ordered and continuous lamellar backbones act like phonon highways and thus support fast propagation of both longitudinal and transversal acoustic waves in the spider dragline silks, as observed in the recent Brillouin scattering measurements \cite{nmat3549}. On the other hand, due to the modulation of the lamellar structure, the crystalline ordering vanishes and the $\beta$-sheet crystal structure has a zero-density discontinuity along the radial direction. Consequently, the scattering of phonons is stronger and the measured velocities of the acoustic wave are much lower than that along the axial direction. 

The other prominent structural feature is the clustered packing of the $\beta$-sheet crystals. During the reconstruction process, $\beta$-sheet crystals self assemble to form high density crystalline-rich islands. The clustering effect exist in all length scale examined here and the cluster sizes change from 30 nm in Fig.\ref{fig:map}c to 50 nm in Fig.\ref{fig:map}b, and to 100 nm in Fig.\ref{fig:map}a, increasing exponentially in accordance with the coarse graining. The reconstructed electron density map shows remarkable self-similarity property and is scale invariance statistically. The formation of these clusters is driven by both the characteristic of the matrix knees and the presence of the lamellar peaks. The SAXS structure factor has power law relation with respect to the scattering vector $q$
\begin{equation}
	S(q) \propto q^{-r}
\end{equation}
and $-r$ is the slope of the matrix knee in the log-log plot of $S(q)$ (Fig.\ref{fig:xray}e). The parameter $r$ is the fractal dimension in system exhibiting self-similarity \cite{Pedersen:gl0327,Teixeira:gk0142,Martin:a27719}. For the silk fibers examined in this study, the $r$ is between 2.16 and 2.68, which means the crystalline structure has a mass fractal ($2<r<3$)\cite{Schaefer1989,BF01012944}. Crystal clustering is the dominant form of mass fractal and therefore the clustering effect can be observed at drastically different length scale. Now looking back at the pair correlation function (Fig.\ref{fig:saxs}b), the correlation peaks beyond 10 nm scale should arise from the inter-cluster interaction. No matter how strong the lamellar peak is, the mass fractal accompanied by clustering is an universal property in spider dragline silk, manifested by the power law of structure factor and the reconstructed electron density map (see Fig.S3). 

The mass fractal property is significant to the mechanical strength of spider silks. The clusters can be viewed as the basic functional mechanical units, and more importantly such functional units exist throughout the intermediate length scale. Whenever the silk experiencing an external force, the largest cluster of the size 100 nm will deform to respond to the external kick, then the deformation is propagated down to the smaller clusters which will deform in response. The force-induced deformation will be propagated down in this manner all the way to the $\beta$-sheets level, causing an exponentially increasing number of deformed functional units as the length scale reduces. As illustrated by Fig.\ref{fig:map}d, the hierarchical self-similar model is remarkably robust. This scheme is consistent with the model proposed by Zhou \& Zhang \cite{PhysRevLett.94.028104} which effectively explained the exponential force-extension property of spider silks \cite{nmat858}. To answer the questions imposed by the large crystalline regions observed in the SEM and TEM images, we suggest that these 20 - 50 nm granules are composed of densely packed, uniformly oriented and strongly interconnected $\beta$-sheet crystals of the sizes 2 to 4 nm.


\section{Experimental methods}
Major ampullate dragline silks were collected by forced silking from living spiders anesthetized with carbon monoxide. The silks were reeled at $2.0\pm0.1$ cm$\cdot$s\textsuperscript{-1} and were directly mounted across hollow cardboard holders driven by an electric motor. At these silking speeds, $\beta$-sheet crystals are supposed to be extremely well oriented \cite{Du20064528}. The samples normally have 50 to 100 strands of fibres.

The X-ray experiment was carried out at Argonne National Laboratory Advanced Photon Source BioCars 14-ID-B beam line optimized with a SAXS setup. The incident beam energy was 9 K\si{\eV}, corresponding to an X-ray wavelength of 1.38 \si{\angstrom}. The scattering data were collected on a Mar165, a 3072 by 3072 pixel resolution CCD detector from Marresearch with a pixel size of \SI{79}{\micro\metre}. The detector to sample distance is fixed at 180 \si{\milli\metre} throughout the experiment. The sample was mounted to an xyz-translation goniometer head with helium gas chamber placed between the CCD detector and the sample to help reduce the background signal. The exposure time varied from 2 \si{\second} to 6 \si{\second} depending on the size of the silk bundle. For each sample ten exposures were taken. Backgrounds were measured after translating the silk bundle out of the X-ray beam. 

The multiple X-ray exposures were averaged and background subtracted by using software \textit{Fit2D} \cite{doi:10.1080/08957959608201408}. The integrated 1-D WAXS and SAXS profile were obtained by using azimuthal integration functionality in \textit{Fit2D}. The automatic Gaussian peak fitting was performed using the \textit{lsqcurvefit} optimization routine from MATLAB. 

\section{Simulation method}
The transformation from electron density map to scattering intensity adopts the fast Fourier transform (FFT) simulation method proposed by Klauss Schmidt-Rohr \cite{Schmidt-Rohr:ce5003}. The electron density map $\rho(\bf{r})$ was represented by an $N\times N$ matrix, where $N$ is usually chosen to be a power of 2, and then converted to a reciprocal space scattering intensity $I(\bf{q})$ by the transformation algorithm. For each sample, we initialized the silk structure such that it conforms to the crystal sizes calculated from the WAXS data and maintained the closest approach distances, which is the d-spacing of SAXS lamellar peak (Tab.1). The scattering centers, which are the $\beta$-sheet crystals in this model, were represented by higher contrast rectangular shape sub-matrix and are uniformly oriented to be parallel to the silk fiber axis. The orientation of the crystals at >2.0 cm$\cdot$s\textsuperscript{-1} can be approximated by perfect alignment with very small error \cite{vanBeek06082002, Du20064528}. A hard-shell exclusion geometry was used to maintain the closest approach distance.

We imposed a model with a lamellar modulation by generating 60 nm wide crystal-rich stripes separated by equal width empty spaces. The lamellar structure is essential to physically represent the silk fibrils fine structure within the silk fiber \cite{kitagawa, Du20064528}.The initial crystalline map was then fed to the stimulated annealing reconstruction routine. The reconstruction was proceeded by generating a electron density map such that the calculated structure factor $\hat{S}(q)$ matches the experimental $S(q)$ with acceptable error tolerance. This was achieved by minimizing the pseudo-energy
\begin{equation}
	E_{pseudo} = \sum_{q}|S(q)-\hat{S}(q)|^{2}
	\label{eq:engy}
\end{equation}
which measures the distance from simulated structure factor to the experimental value \cite{doi:10.1080/08927028808080958,PhysRev.168.1068,Jiao20102009}. The optimization was realized by the stimulated annealing algorithm where each random walk is accepted or rejected by a probability of  \cite{Kirkpatrick13051983}
\begin{equation}
	P(s' \leftarrow s) = min(1,exp(-\frac{\Delta E_{s' \leftarrow s}}{k_{B}T}))
	\label{eq:prob}
\end{equation}
where $s$ and $s'$ are the states before and after one random walk, $\Delta E_{s' \leftarrow s}$ is the change of pseudo-energy after accepted state transition, $k_{B}$ is the Boltzmann constant, $T$ is the imaginary stimulated annealing temperature.

\section{Acknowlegments}
This work was supported by the Department of Defense, AFOSR (FA9550-14-1-0014) and the US National Science Foundation (DMR-1264801). The author would like to thank Klaus Schmidt-Rohr for providing the MATLAB simulation code, Robert Henning at APS BioCars for setting up 14-ID-B beamline and Yang Jiao for instructive discussion. This research used resources of the Advanced Photon Source, a U.S. Department of Energy (DOE) Office of Science User Facility operated for the DOE Office of Science by Argonne National Laboratory under Contract No. DE-AC02-06CH11357. Use of the BioCARS Sector 14 was supported by the National Institutes of Health, National Center for Research Resources, under grant number RR007707. 

\bibliographystyle{apsrev4-1} 
\bibliography{saxs}

\end{document}